# Sum rules in zone axis STEM-orbital angular momentum resolved electron magnetic chiral dichroism

Matteo Zanfrognini[1], Enzo Rotunno[1], Jan Rusz[2], Rafal E. Dunin Borkowski[3], Ebrahim Karimi[4], Stefano Frabboni[1,5] and Vincenzo Grillo[1]

[1]CNR-NANO Via G. Campi 213/a, I-41125 Modena, Italy
[2]Department of Physics and Astronomy, Uppsala University, Box 530, S-751 21 Uppsala, Sweden
[3]Ernst Ruska-Centre for Microscopy and Spectroscopy with Electrons and Peter Grünberg Institute, Forschungszentrum Jülich, 52425 Jülich, Germany
[4]Department of Physics, University of Ottawa, 150 Louis Pasteur, Ottawa, Ontario K1N 6N5, Canada
[5]Dipartimento FIM, Università di Modena e Reggio Emilia, Via G. Campi 213/a, I-41125 Modena, Italy

In this work we derive sum rules for orbital angular momentum(OAM) resolved electron magnetic chiral dichroism (EMCD) which enable the evaluation of the strength of spin and orbital components of the atomic magnetic moments in a crystalline sample. We also demonstrate through numerical simulations that these rules appear to be only slightly dependent from the dynamical diffraction of the electron beam in the sample, making possible their application without the need of additional dynamical diffraction calculations.

## I. INTRODUCTION

Since the work of Schattschneider [1] and collaborators, electron magnetic circular dichroism (EMCD) has stimulated the attention of many researchers in the field of electron microscopy because of its potential of providing information about magnetic properties of materials with sub-nanometric resolution. As in the case of X ray circular dichroism (XMCD)[2,3] sum rules for EMCD, independently derived by Calmels[4] and Rusz[5] permit in principle to quantify the orbital and the spin components of the magnetic moment per atom in the sample, even if a practical application of these rules is made complicated by dynamical diffraction effects, which introduce thickness dependent factors to be evaluated by separate dynamical calculations.

In this work we derive sum rules for the orbital and spin components of the magnetic moments of the atoms in a crystalline sample for the specific case of zone axis orbital angular momentum (OAM) resolved STEM-EMCD, a technique recently proposed in Ref. 6: in such a proposed experiment, both the energy and the OAM spectra[7] of the electrons, having experienced a core-loss process, are measured and differences among the $\ell=\pm 1$ spectra are expected in the case of magnetic materials. The main result of this analysis is that, if the probe evolution along the column is dominated by channeling ( as in most zone-axis STEM experiments), the OAM resolved EMCD sum rules are strongly simplified: practically, they exhibit a weak dependency from diffraction effects, making their application more straightforward in real life experiments.

## II. THEORY

The experimental quantity which can be directly evaluated through the combined action of an OAM sorter[7] and an energy spectrometer in TEM is the OAM-resolved loss function $I_\beta(\ell, \Delta E)$ which can be formally defined as[6,8]

$$I_\beta(\ell, \Delta E) = \int_0^{\frac{2\pi\beta}{\lambda}} I(\ell, k, \Delta E) dk = \int_0^{\frac{2\pi\beta}{\lambda}} Tr[\hat{\rho}_f \hat{P}_{\ell E}(k)] dk \quad (1)$$

in which $\hat{\rho}_f$ is the density matrix of the electrons inelastically scattered by the sample and $\hat{P}_{\ell E}(k)$ is a unitary projector over the states $|\ell, k, \Delta E>$ having OAM equal to $\hbar\ell$ along axis $z$ (the TEM optical axis) and energy $E = E_0 - \Delta E$, being $E_0$ the energy of the incoming electron beam; $k$ corresponds to the electronic scattering wave vector in the $xy$ plane, $\lambda$ is the electronic de Broglie wavelength and $\beta$ is the semi-collection angle of the OAM spectrometer, over which the electrons are collected after the sample. With these definitions, $k\lambda/2\pi$ represents scattering angle (given below in mrad units).

Assuming valid dipolar approximation[6], taking the material magnetization saturated along $z$ axis and working in paraxial conditions for the incoming electrons, it is possible to write $I(\ell, k, \Delta E)$ appearing in Eq.(1) as[6,9]

$$I(\ell, k, \Delta E) = \sum_{i,j}^{x,y,z} N_{ij}(\Delta E) \sum_a P_a^{ij}(\ell, k) + M(\Delta E) \sum_a S_a^z(\ell, k) \quad (2)$$

where $N_{ij}(\Delta E)$ and $M(\Delta E)$ only depend on the electronic properties of the material and respectively describe the non magnetic and the magnetic contributions to the loss spectrum. The functions $P_a^{ij}(\ell, k)$ and $S_a^z(\ell, k)$ define the effects of the dynamical diffraction of the electron beam in the sample: the summations over the atomic positions $a$ are restricted over the magnetic atoms which can give rise to the energy losses $\Delta E$ in which we are interested; we point out here that Eq.(2) is rigorously valid only if the magnetic atom of interest in the whole sample are equivalent one with the others, otherwise a dependency on $a$ should also appear for $N_{ij}(\Delta E)$ and $M(\Delta E)$.

Integrating Eq.(2) over the scattering angle $k\lambda/2\pi$, we can write the OAM resolved loss function as

$$I_\beta(\ell, \Delta E) = \sum_{i,j}^{x,y,z} N_{ij}(\Delta E) \sum_a F_a^{ij}(\ell, \beta) + M(\Delta E) \sum_a S_a^z(\ell, \beta) \quad (3)$$

being
$$F_a^{ij}(\ell,\beta) = \int_0^{\frac{2\pi\beta}{\lambda}} P_a^{ij}(\ell,k)dk$$

and

$$S_a^z(\ell,\beta) = \int_0^{\frac{2\pi\beta}{\lambda}} S_z(\ell,k)dk$$

In conventional STEM experiments, once the beam is centered on an atomic column, the electrons tend to be laterally confined along it, in a phenomenon called channeling[10,11,12]: practically, the probing electrons form a laterally bound state in the projected potential exerted by the atoms so only a small fraction of the beam can go away from it (i.e. can de-channel), while propagating in the sample.

As core loss scattering is a strongly localized process, the main contributions to the overall inelastic signal come from the atoms of the column on which the STEM probe is focused, while those of the neighboring columns only partially contribute to such a quantity, as only a small portion of the incoming electrons de-channel towards these atoms: therefore one should expect that all the dynamical coefficients $P_a^{ij}$ and $S_a^z$ have an intense strength for $a$ on the column on which the probe is centered ( from now on this column will be taken at (0,0) in the $xy$ plane) and a rapidly decreasing value for the atoms on the neighboring and more distant ones.

We now take into account the role of the dispersion in OAM performed through appropriate devices introduced in the TEM column[7]: practically, we focus on the features of the OAM dispersed coefficients $P_a^{ij}(\ell,k)$ and their integrated counterparts $F_a^{ij}(\ell,\beta)$ evaluated for $\ell = \pm 1$, respectively as functions of $k$ and $\beta$. As we formally demonstrate in Appendix A, some of these quantities are zero by symmetry once evaluated for the atoms along the central column with coordinates $(0,0,a_z)$ and they are expected to be non-zero only for the neighboring columns where, in optimal channeling conditions, the probe intensity is low, making negligible their contribution to the function $I_\beta(\ell, \Delta E)$.

The main results derived in Appendix A can be summarized by the following formulas

$P_{(0,0,a_z)}^{iz}(\ell = \pm 1,k) = 0, \forall k, \text{with } i = x,y,z$ (4.1)
$P_{(0,0,a_z)}^{xy}(\ell = \pm 1,k) = 0, \forall k$ (4.2)
$P_{(0,0,a_z)}^{xx}(\ell = \pm 1,k) = P_{(0,0,a_z)}^{yy}(\ell = \pm 1,k) \forall k$ (4.3)

Such relations, together with the fact that the contributions coming from the neighboring columns are negligible, suggest that the functions $P_a^{xx}(\ell,k)$ and $P_a^{yy}(\ell,k)$ (and so $F_a^{xx}(\ell,\beta)$ and $F_a^{yy}(\ell,\beta)$) dominate the non magnetic part of the OAM resolved loss function, computed for $\ell = \pm 1$. These rather general predictions are here confirmed and detailed in a few specific material systems of interest. Practically, we will describe the behavior of the summed quantities $F^{ij}(\ell = +1,\beta) = \sum_a F_a^{ij}(\ell,\beta)$ as functions of the semi-collection angle $\beta$ for the specific case of a Cobalt sample of 30 nm, oriented along the [001] zone axis, keeping into account the contributions of all the magnetic atoms in the sample.

Further, in the Supplementary Material (SM), we also show the results of similar calculations performed for bcc Iron and FePt samples, in order to demonstrate the quite general validity of the considerations exposed above.

The calculation of the functions $P^{ij}(\ell = +1,k)$ (and so of the integrated counterparts $F_a^{ij}(\ell,\beta)$) has been performed by a multislice approach through a modified version of the software MATSv2[13], according to the procedure outlined in Refs.[6][9][13]: details about the convergence parameters and the unit cells adopted to perform the simulations of the electron beam propagation in the crystal are given in the SM.

In all calculations presented in this work, the incident electron beam is assumed with energy of 300 keV and different semi-convergence angles, in a range from 7 to 22 mrad, to explore different channeling conditions for the STEM probe.

Calculations for $\ell = -1$ are presented in the SM: from those results it is simple to realize that the considerations exposed here for $\ell = +1$ can be easily generalized to the opposite OAM.

The results of these calculations are summarized in Figures 1 and 2. Looking at Fig. 1, it is simple to observe that the behavior of the functions $F^{xx}(\ell = +1,\beta)$ with $\beta$ turns out to be practically unmodified by modifying the STEM probe convergence within reasonable limits: the increase in its absolute value obtained increasing the probe convergence from 7 mrad to 16 mrad can be justified by the fact that the latter beam channels in a stronger way along the Co atomic columns than the former, and so an overall increase in the inelastic signal with $\ell = +1$ is expected. Further, looking at the inset a) of Fig.1 we point out that the ratio $\frac{P^{xx}(\ell=+1,k)}{P^{yy}(\ell=+1,k)}$ turns out to be almost constant with the scattering angle $k$, with very small fluctuations around one, as predicted by Eq. 4.3, considering only the atoms on the (0,0) column.

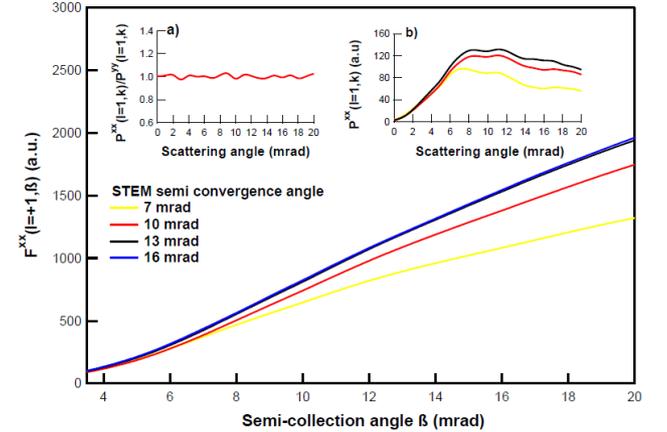

**Figure 1** Function $F^{xx}(\ell = 1,\beta)$ computed for STEM probes with different semi-convergence angle (from 7 mrad to 16 mrad) in the case of a hpc Cobalt sample of 30 nm oriented along the [001] direction: the behavior of this function with $\beta$ turns out to be practically unmodified by changing the STEM probe convergence; the increase in its absolute value obtained passing from the 7 mrad probe to the one with convergence of 16 mrad is justified by the fact that the latter beam channels in a stronger way along the Co atomic columns than the former. In the inset a) the ratio $\frac{P^{xx}(\ell=+1,k)}{P^{yy}(\ell=+1,k)}$ is presented (evaluated for a probe semi-convergence of 10 mrad): it is simple to notice that this quantity only weakly oscillates around one as a function of the scattering angle; finally, in inset b) the function $P^{xx}(\ell = +1,k)$ is shown for different STEM convergence angles (see main legend).

The small differences we notice in the inset a) of Fig. 1 are only due to the atoms of the neighboring columns, whose contributions to the overall inelastic signal are much smaller than the ones from the column in (0,0).

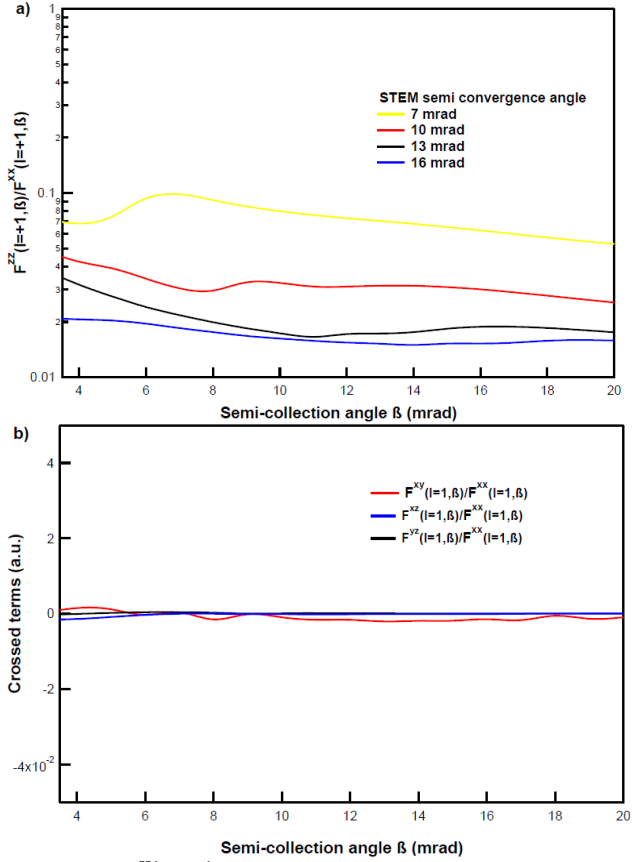

**Figure2** a) Ratio $\frac{F^{zz}(\ell=+1,\beta)}{F^{xx}(\ell=+1,\beta)}$ for different STEM semi convergence angles (from 7 mrad to 16 mrad): notice that the function $F^{zz}(\ell=+1,\beta)$ is almost two orders of magnitude smaller than $F^{xx}(\ell=+1,\beta)$, and such a ratio decreases by increasing the probe convergence, in this range. b) Functions $\frac{F^{xy}(\ell=+1,\beta)}{F^{xx}(\ell=+1,\beta)}$, $\frac{F^{xz}(\ell=+1,\beta)}{F^{xx}(\ell=+1,\beta)}$ and $\frac{F^{yz}(\ell=+1,\beta)}{F^{xx}(\ell=+1,\beta)}$ computed assuming a STEM probe of 10 mrad: it is simple to realize that the cross terms appearing in Eq. 3 are effectively at least two orders of magnitude smaller than $F^{xx}(\ell=+1,\beta)$, independently from the semi-collection angle $\beta$.

In Figure 2a we show $\frac{F^{zz}(\ell=+1)}{F^{xx}(\ell=+1)}$ (in log scale) as a function of the semi-collection angle $\beta$, computed for different STEM probe convergences. This ratio points out that $F^{xx}(\ell=+1,\beta)$ is at least an order of magnitude larger than $F^{zz}(\ell=+1,\beta)$, and such a ratio decreases by increasing the probe convergences, i.e. the channeling capability of the incoming electron beam: this can be justified remembering that the contribution of the atoms $(0,0,a_z)$ to the function $P^{zz}(\ell=+1,k)$ (and so to $F^{zz}(\ell=+1,\beta)$) is zero by symmetry as clarified by Eq. 4.1.

Finally, in Figure 2b the cross terms $F^{ij}(\ell=+1,\beta)$ are shown, for a probe convergence of 10 mrad: it is evident that these functions are all at least two orders of magnitude smaller than $F^{xx}(\ell=+1,k)$ and $F^{yy}(\ell=+1,k)$; again this behavior is related to the fact that the atoms on the column on which the beam is focused do not give contribution to these functions, so that their value turns out to be effectively negligible with respect to the one of $F^{ii}$, with $i=x,y$.

These calculations permit to simplify Eq.3, through the following reasonable assumptions:

- we neglect the contributions to $I_\beta(\ell=\pm1,\Delta E)$ due both to the cross terms $F^{ij}(\ell=\pm1,\beta)$ (with $i\neq j$) and to $F^{zz}(\ell=\pm1,\beta)$, as they are at least two orders of magnitude smaller than $F^{ii}(\ell=\pm1,\beta)$, with $i=x,y$;
- as the ratio $\frac{P^{xx}(\ell=+1,k)}{P^{yy}(\ell=+1,k)}$ only slightly fluctuates around one, we take $P^{yy}(\ell=+1,k)\approx P^{xx}(\ell=+1,k)$, and so $F^{xx}(\ell=\pm1,\beta)\approx F^{yy}(\ell=\pm1,\beta)=F_\beta(\ell=\pm1)$.

We underline that these conclusions have general validity, independently from the chosen material and its symmetry: the only requirements are orienting the crystal along an high symmetry direction and using an electron probe characterized by strong channeling properties along the chosen atomic column.

Under these approximations, we can write the inelastic signal experimentally observed at energy $\Delta E$ and at OAM $\ell\hbar=\pm\hbar$ as

$$I_\beta(\ell=\pm1,\Delta E) \approx [N_{xx}(\Delta E)+N_{yy}(\Delta E)]F_\beta(\ell=\pm1) + M(\Delta E)S^z(\ell=\pm1,\beta) \quad (5)$$

which will be used in the following section to derive the sum rules for OAM resolved EMCD.

## III. DERIVATION OF SUM RULES FOR OAM-RESOLVED EMCD

The aim of this section is to derive sum rules for OAM resolved EELS experiment, starting from the expression of $I_\beta(\ell=\pm1,\Delta E)$ obtained in Section II: practically we will derive expressions of the orbital and spin components of the atomic magnetic moment as functions of the measured OAM resolved EEL spectra, directly available experimentally.

To do this we will relate the functions $N_{ii}(\Delta E)$ and $M(\Delta E)$ to the mixed dynamic form factor (MDFF)[14] exploiting the rules provided in Ref. 5 and reported for clarity in Section 1 of the SM. If the sample, in the selected projection, is mirror symmetric with respect to a plane perpendicular to $xy$ plane, we have demonstrated in Ref.6 that

$$P^{xx}(\ell,k) = P^{xx}(-\ell,k)$$
$$S^z(-\ell,k) = -S^z(\ell,k)$$

where we have assumed to define the $x$ axis as the intersection of this symmetry plane with the one perpendicular to $z$.

Using these properties it is simple to write

$$I_\beta(+1,\Delta E)+I_\beta(-1,\Delta E) = 2[N_{xx}(\Delta E)+N_{yy}(\Delta E)]F_\beta(+1) \quad (6.1)$$

$$I_\beta(+1,\Delta E)-I_\beta(-1,\Delta E) = 2M(\Delta E)|S^z(+1,\beta)| \quad (6.2)$$

Assuming to work with *3d* transition metals, we provide the following definitions

$$A_2 = \int_{L_2} d\Delta E\,[I_\beta(+1,\Delta E)-I_\beta(-1,\Delta E)] \quad (7.1)$$

$$A_3 = \int_{L_3} d\Delta E\,[I_\beta(+1,\Delta E)-I_\beta(-1,\Delta E)] \quad (7.2)$$

$$A_{23} = \int_{L_2+L_3} d\Delta E\,[I_\beta(+1,\Delta E)+I_\beta(-1,\Delta E)] \quad (7.3)$$

$A_{23}$ corresponds to the integral of the sum of the $\ell=\pm1$ spectra over the atomic edges $L_2$ and $L_3$, while $A_2$ ($A_3$) is the integral over the single edge $L_2$ ($L_3$) of their difference. We now use these quantities to obtain sum rules for the orbital and the spin components of the atomic magnetic moment.

**Orbital sum rule**

Using Eq. 6 and 7 we can write

$$\frac{A_2 + A_3}{A_{23}} = \frac{\int_{L_2+L_3} M(\Delta E) d\Delta E}{\int_{L_2+L_3} [N_{xx}(\Delta E) + N_{yy}(\Delta E)] d\Delta E} \frac{|S^z(1,\beta)|}{F_\beta(+1)} \quad (8)$$

As demonstrated in Appendix B, we have

$$\int_{L_2+L_3} [N_{xx}(\Delta E) + N_{yy}(\Delta E)] d\Delta E = \frac{9L^2 G_L^2}{(2L-1)(2L+1)}[D_{xx} + D_{yy}]$$

and

$$\int_{L_2+L_3} M(\Delta E) d\Delta E = \frac{9 G_L^2}{2\mu_B(2L+1)} m_{orb}$$

where $L = 2$, $m_{orb}$ is the orbital component of the atomic magnetic moment and the functions $G_L^2$, $D_{xx}$ and $D_{yy}$ depend only on the material electronic properties and can be determined through *ab initio* calculations.

By simple substitution of these expressions in Eq. (8) we can immediately find a sum rule for the orbital component $m_{orb}$ given by

$$m_{orb} = \frac{2\mu_B L^2 [D_{xx} + D_{yy}]}{2L-1} \frac{F_\beta(+1)}{|S^z(1,\beta)|} \frac{A_2 + A_3}{A_{23}} \quad (9)$$

**Spin sum rule**

Using Eqs. 6.1 and 6.2, we have

$$\frac{1}{L-1} A_2 = \frac{2M_2}{L-1} |S^z(+1,\beta)|$$
$$\frac{1}{L} A_3 = \frac{2M_3}{L} |S^z(+1,\beta)|$$

where $M_i$ stands for the integration of $M(\Delta E)$ over the atomic edge $L_i$: these quantities are evaluated in Appendix B.

Exploiting the analytical expressions of $M_2$ and $M_3$ we can immediately find

$$\frac{1}{L} A_3 - \frac{1}{L-1} A_2 = \frac{18 |S^z(+1,\beta)| G_L^2}{(2L-1)(2L+1)} \frac{2L-1}{3\mu_B} m_{spin}$$

where $m_{spin}$ is the spin component of the atomic magnetic moment. Dividing this difference for the integral over the two edges of the sum of $I_\beta(+1, \Delta E)$ and $I_\beta(-1, \Delta E)$ we find

$$m_{spin} = \frac{3\mu_B L^2 [D_{xx} + D_{yy}]}{2L-1} \frac{F_\beta(+1)}{|S^z(+1,\beta)|} \frac{\frac{1}{L} A_3 - \frac{1}{L-1} A_2}{A_{23}} \quad (10)$$

which corresponds to the desired sum rule for the spin component of the atomic magnetic moment.

By computing the ratio $\frac{m_{orb}}{m_{spin}}$ using Eqs. 9 and 10 we find

$$\frac{m_{orb}}{m_{spin}} = \frac{2}{3} \left[ \frac{A_2 + A_3}{\frac{1}{L} A_3 - \frac{1}{L-1} A_2} \right]$$

i.e. the ratio of the orbital and spin components of the atomic magnetic moment does not depend on dynamical effects and *ab initio* pre-factors, and so can be evaluated directly from the experimental spectra: such relation was first reported in Ref.[4].

## IV. DEPENDENCY OF SUM RULES FROM DYNAMICAL DIFFRACTION EFFECTS

Looking at Eq. 9 and Eq. 10 we notice both the presence of terms which need to be evaluated *ab initio* and also of a quantity which only depends on the dynamical diffraction of the electron beam in the sample; in the following we define this factor as $R_\beta$, given by

$$R_\beta = \frac{F_\beta(+1)}{|S^z(+1,\beta)|}$$

Our simulations point out that this ratio is in reality almost independent from the crystal under study, the sample thickness and, for sufficiently large $\beta$, from the OAM spectrometer semi-collection angle. This can be understood from Figure 3, where such a quantity is evaluated as a function of $\beta$ for different convergences of the STEM probe for a Cobalt sample of 30 nm. We notice that choosing $\beta \geq 8$ mrad, $R_\beta$ converges (from above) to a value close to 0.5: as shown in the SM, this behavior is also observed for a bcc Iron and a FePt sample with thicknesses of 20 nm.

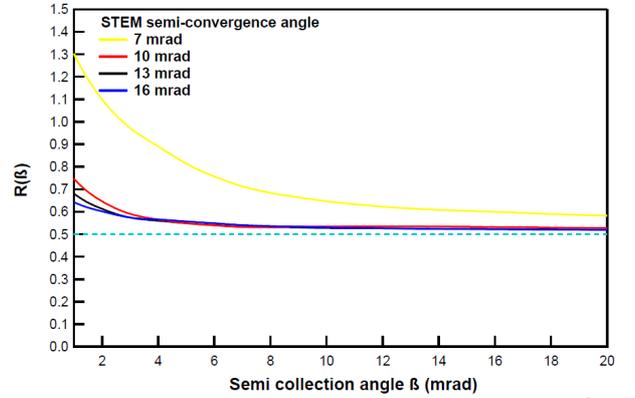

**Figure 3** Function **R** computed for different semi-collection angles $\beta$ and for different STEM convergences. Notice that for a small convergence (i.e. 7 mrad) this function converges to a value quite larger than 0.5, because of the strong de-channelling suffered by this electron beam (see SM for more details). The dashed line corresponds to the result we would obtain for a perfectly channeling beam (like a gaussian beam), which should be able to propagate in the crystal without exciting atoms not on the column on which it is centered.

Such a general behavior can be understood from the calculations we present in Appendix A: practically, we find that once $\mathbf{a} = (0,0,a_z)$, the following relation holds

$$|S_a^z(\ell = \pm 1, k)| = 2P_a^{xx}(\ell = \pm 1, k)$$

Therefore using the definitions of $F_\beta$ and $S^z(+1,\beta)$ and neglecting the contributions of the atoms not on the (0,0) column we have $R_\beta^{(0,0,a_z)} \rightarrow 1/2$. This asymptotic behavior could be confirmed taking as a probe a Gaussian beam with transverse size matching the one of the Bloch 1s state of the Cobalt crystal: such a probe[15] exhibits almost perfect channeling along the atomic column on which the beam is centered i.e. it is able to propagate in the crystal almost without any diffraction effect, similar to Bessel beams in vacuum[16]. Because of this property, such a beam is expected not to excite the atoms of the other columns so that we should provide $R_\beta \rightarrow 1/2$.

The fact that $R_\beta$ never converges to this limiting value for conventional probes is only due to the presence of the neighboring atomic columns; to further confirm this point, we have evaluated $R_\beta$ with different probe convergences and we have collected the values assumed by these ratio for $\beta = 10$ mrad, i.e. when $R_\beta$ starts assuming a flat trend, as a function of the semi-collection angle: the results are shown in Figure 4. We notice that by passing from a semi-convergence angle of 7 mrad to one of 13mrad, $R_\beta$ gets closer to the limiting value of 0.5, while for larger angles it stays almost constant: this is due to the fact that for the latter probes the degree of de-channelling is strongly reduced, with respect to the former one; this last point is further clarified through separate multislice calculations presented in the SM.

Therefore, from our results, it emerges that, by appropriately choosing the STEM convergence and the collection angle $\beta$ so to have $R_\beta$ as much as possible close to 0.5, sum rules for OAM resolved EMCD turn out to be only weakly dependent from dynamical diffraction effects.

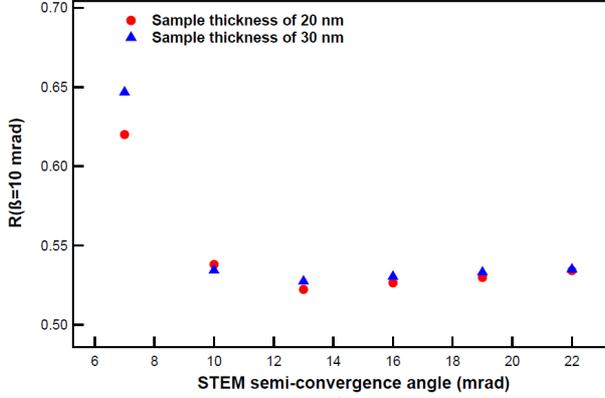

**Figure 4** Function $R_\beta$ evaluated for $\beta = 10$ mrad for different semi-convergence angles of the STEM probe: the calculations have been performed for a Cobalt sample of 20 nm (red circles) and one of 30 nm (blue triangles): notice that in both cases for a STEM convergence of about 13 mrad $R_\beta \approx 0.52$, which corresponds to an overestimation of the components of the magnetic moment of about 4 %.

To quantify the error in the evaluation of $m_{orb}$ (but also of $m_{spin}$) once $R$ is approximated by 0.5, we call $R = 0.5 + \delta R$, where $\delta R$ is the deviation from the limiting value only due to the atoms on the column $(0,0)$. Thus the measured orbital contribution $m_{orb}^{exp}$ will be

$$m_{orb}^{exp} = m_{orb}^{exact} + 2m_{orb}^{exact}\delta R$$

where $m_{orb}^{exact}$ is the real orbital component. Therefore, the experimental orbital component overestimates the real one by a fraction $\Delta$ given by

$$\Delta = \frac{m_{orb}^{exp} - m_{orb}^{exact}}{m_{orb}^{exact}} = 2\delta R$$

that depends from sample thickness and geometry, but it is expected to be typically within the experimental error: in the simulations performed here, $\delta R$ is always smaller than 0.04 (for convergence angles larger than 7 mrad), which corresponds to an error (at maximum) of the order of 7-8%, which can be reduced to about 4% using a 13 mrad electron beam.

## V. CONCLUSIONS

In this work we have derived sum rules for OAM resolved EMCD and we have demonstrated how these relations are characterized by a weak dependence from the dynamical diffraction effects, which can also be minimized by an appropriate choice of the STEM probe to be used in experiments. These results should make the application of sum rules easier from the experimental point of view and, combined with the atomic resolution provided by OAM resolved EMCD, they could give access to atomically resolved maps for the orbital and spin components of the atomic magnetic moments in crystalline samples.

## ACKNOWLEDGEMENTS

This work is supported by Q-SORT, a project funded by the European Union's Horizon 2020 Research and Innovation Program under grant agreement No. 766970.

## Appendix A

In analogy with Refs. [6][13], we introduce the following definitions for the OAM resolved dynamical coefficients appearing in Eq. (2)

- $P^{ii}(\ell, k) = \sum_a |Q_a^i(\ell, k)|^2 = \sum_a P_a^{ii}(\ell, k)$
- $P^{ij}(\ell, k) = 2\sum_a Re[Q_a^i(\ell, k)Q_a^j(\ell, k)^*] = \sum_a P_a^{ij}(\ell, k)$
- $S^z(\ell, k) = -2\sum_a Im[Q_a^x(\ell, k)Q_a^y(\ell, k)^*] = \sum_a S_a^z(\ell, k)$

for $i$ and $j = x, y, z$, where $Q_a^i(\ell, k)$ are auxiliary functions defined in Refs. [6][9][13].

In this section we provide a description of the properties of $Q_a^i(\ell, k)$ for $\boldsymbol{a} = (0, 0, a_z)$, i.e. for the magnetic atoms positioned on the column on which the STEM probe is focused.

We start defining $Q_{(0,0,a_z)}^i(\ell, k) = Q_a^i(\ell, k)$ as

$$Q_a^i(\ell, k) = \sigma \int d\boldsymbol{k}_1 \int d\boldsymbol{k}_2 D^*(\boldsymbol{k}_1; \ell, k) C(\boldsymbol{k}_2) \frac{\tilde{q}_i}{\tilde{q}^2} e^{i(k_1^z - k_2^z)a} \quad (A1)$$

where $\tilde{\boldsymbol{q}} = \boldsymbol{k}_1^\perp - \boldsymbol{k}_2^\perp + q_{\Delta E}\boldsymbol{z}$, with $q_{\Delta E}$ defined as in Eq.5 of Ref.9, while $\sigma$ is a constant dependent on the energy of the probing electrons. Following [6][9] we have that

$$C(\boldsymbol{k}) = \int d\boldsymbol{x} e^{-i\boldsymbol{k}\boldsymbol{x}}[\hat{U}(z, 0; \boldsymbol{x}^\perp)\phi_{inc}(\boldsymbol{x}^\perp; z = 0)] \quad (A2)$$

$$D^*(\boldsymbol{k}; \ell, k) = \int d\boldsymbol{x} e^{i\boldsymbol{k}\boldsymbol{x}}[\hat{U}^+(z, 0; \boldsymbol{x}^\perp)J_\ell(kr)e^{i\ell\varphi}]^* \quad (A3)$$

From Eq.(A2) and Eq.(A3) it is possible to understand that $C(\boldsymbol{k})$ and $D^*(\boldsymbol{k}; \ell, k)$ are respectively the 3D Fourier transform of the incoming beam propagating inside the crystal and of a Bessel beam (with topological charge $\ell$ and transverse wavevector $k$) back-propagating in the sample: in fact $\hat{U}(z, 0; \boldsymbol{x}^\perp)$ is an operator (defined in Eq.(3) of Ref. 9) which provides the wavefunction in the crystal at $z$, in position $\boldsymbol{x}^\perp$ in a plane perpendicular to the TEM optical axis, given the wavefunction at the same point $\boldsymbol{x}^\perp$ at plane $z= 0$, i.e. the entrance plane of the sample.

We now assume the crystal to be invariant under rotation of an angle $\alpha$ in the interval $(0, \pi]$ around an axis parallel to $z$ and centered in $(0;0)$: this hypothesis is satisfied every time the crystal is oriented along an high symmetry direction. Let's start from the expression of $Q_a^x(\ell, k)$ and consider the following substitutions
$$k_1^{x\prime} = k_1^x \cos\alpha - k_1^y \sin\alpha$$
$$k_1^{y\prime} = k_1^x \sin\alpha + k_1^y \cos\alpha$$
$$k_2^{x\prime} = k_2^x \cos\alpha - k_2^y \sin\alpha$$
$$k_2^{y\prime} = k_2^x \sin\alpha + k_2^y \cos\alpha$$
Therefore, $k_1$ is a function of $k'_1$ and $k_2$ depends on $k'_2$, i.e. $k_1 = k_1(k'_1)$ and $k_2 = k_2(k'_2)$. Remembering that the Jacobian of this transformation is one, we can write
$$Q_a^x(\ell, k) = \sigma \int dk'_1 \int dk'_2 D^*(k_1(k'_1); \ell, k) C(k_2(k'_2))$$
$$\times \frac{\left(k_1^{x\prime} - k_2^{x\prime}\right)\cos\alpha + \left(k_1^{y\prime} - k_2^{y\prime}\right)\sin\alpha}{|k_1^{\perp\prime} - k_2^{\perp\prime}|^2 + q_{\Delta E}^2} e^{i(k_1^z - k_2^z)a} \quad (A4)$$
where
$$D^*(k_1(k'_1); \ell, k) = \int dz \int dx^\perp e^{i(k_1^{x\prime}\cos\alpha + k_1^{y\prime}\sin\alpha)x} e^{i(-k_1^{x\prime}\sin\alpha + k_1^{y\prime}\cos\alpha)y}$$
$$\times [\hat{U}^+(z, 0; x^\perp) J_\ell(kr) e^{i\ell\varphi}]^*$$
Calling
$$x' = x\cos\alpha - y\sin\alpha$$
$$y' = x\sin\alpha + y\cos\alpha$$
and reminding $\hat{U}^+(z, 0; x(x', y'), y(x', y')) = \hat{U}^+(z, 0; x', y')$ as this evolution operator depends on the crystal potential which is, by hypothesis, invariant under the transformation $(x, y) \to (x', y')$, we have
$$D^*(k_1(k'_1); \ell, k) = \int dz \int dx'^\perp e^{ik_1^{x\prime}x'} e^{ik_1^{y\prime}y'}$$
$$\times [\hat{U}^+(z, 0; x'^\perp) J_\ell(kr) e^{i\ell\varphi(x', y')}]^*$$
where
$$\tan\varphi(x', y') = \frac{y(x', y')}{x(x', y')} = \frac{y' - x'\tan\alpha}{x' + y'\tan\alpha} = \frac{\frac{y'}{x'} - \tan\alpha}{1 + \frac{y'}{x'}\tan\alpha}$$
$$= \tan(\varphi' - \alpha)$$
from which $\varphi(x', y') = \varphi' - \alpha$. So it is simple to realize
$$D^*(k_1(k'_1); \ell, k) = e^{i\ell\alpha} D^*(k'_1; \ell, k) \quad (A5)$$
Proceeding in the same way, it is possible to demonstrate that
$$C(k_2(k'_2)) = C(k'_2) \quad (A6)$$
By substitution of Eq.(A5) and Eq.(A6) in Eq.(A4) we obtain
$$Q_a^x(\ell, k) = \sigma e^{i\ell\alpha} \int dk'_1 \int dk'_2 D^*(k'_1; \ell, k) C(k'_2)$$
$$\times \frac{\left(k_1^{x\prime} - k_2^{x\prime}\right)\cos\alpha + \left(k_1^{y\prime} - k_2^{y\prime}\right)\sin\alpha}{|k_1^{\perp\prime} - k_2^{\perp\prime}|^2 + q_{\Delta E}^2} e^{i(k_1^z - k_2^z)a}$$
which can be written as
$$Q_a^x(\ell, k) = e^{i\ell\alpha}[Q_a^x(\ell, k)\cos\alpha + Q_a^y(\ell, k)\sin\alpha]$$
from which a relation among $Q_a^x(\ell, k)$ and $Q_a^y(\ell, k)$ can be obtained, i.e.
$$Q_a^x(\ell, k) = \frac{e^{i\ell\alpha}\sin\alpha}{1 - e^{i\ell\alpha}\cos\alpha} Q_a^y(\ell, k) \quad (A7)$$
where the pre-factor is equal to $i/\ell$ once $\alpha$ tends to $\pi$.
If we now consider the definition of $Q_a^z(\ell, k)$, i.e.
$$Q_a^z(\ell, k) = \sigma \int dk_1 \int dk_2 D^*(k_1; \ell, k) C(k_2) \frac{q_{\Delta E}}{\tilde{q}^2} e^{i(k_1^z - k_2^z)a}$$
and we perform the same precedent substitutions we can find
$$Q_a^z(\ell, k) = \sigma e^{i\ell\alpha} \int dk'_1 \int dk'_2 D^*(k'_1; \ell, k) C(k'_2) \frac{q_{\Delta E} e^{i(k_1^z - k_2^z)a}}{|k_1^{\perp\prime} - k_2^{\perp\prime}|^2 + q_{\Delta E}^2}$$
i.e.
$$Q_a^z(\ell, k) = e^{i\ell\alpha} Q_a^z(\ell, k) \quad (A8)$$
Exploiting Eq. (A7) and (A8), we now demonstrate that only the functions $P_a^{xx}(\ell = \pm 1, k)$, $P_a^{yy}(\ell = \pm 1, k)$ and $S_a^z(\ell = \pm 1, k)$ are not zero if evaluated for $a = (0, 0, a_z)$; at the same time we also find relations among these non zero terms.

Starting from Eq. (A8), it is simple to realize that $Q_a^z(\ell, k)$ is not zero only if $(1 - e^{i\ell\alpha}) = 0$. If $\ell = \pm 1$, this means that $Q_a^z(\ell, k)$ is non zero only if the rotation angle for which the crystal is invariant is an integer multiple of $2\pi$, which are, by hypothesis, neglected in the present treatment: so $Q_a^z(\ell = \pm 1, k) = 0$ and $P_a^{iz}(\ell = \pm 1, k) = 0$, for $i = x, y, z$.

If we now take into account Eq.(A7), we can write
$$P_a^{xx}(\ell, k) = \frac{(\sin\alpha)^2}{(1 - e^{i\ell\alpha}\cos\alpha)(1 - e^{-i\ell\alpha}\cos\alpha)} P_a^{yy}(\ell, k)$$
which becomes, for $\ell = \pm 1$,
$$P_a^{xx}(\ell = \pm 1, k) = P_a^{yy}(\ell = \pm 1, k) \quad (A9)$$
so the contribution to $P^{xx}$ due to the atoms on the column on which the STEM probe is centered equals the contribution to $P^{yy}$ for $\ell = \pm 1$.

Finally, using the definitions of $Q_a^i(\ell, k)$ given in Eq.A1, we can compute the quantity
$$2Q_a^x(\ell, k)Q_a^y(\ell, k)^* = P_a^{xy}(\ell, k) - iS_z^a(\ell, k)$$
Exploiting Eq. A7, we obtain
$$2Q_a^x(\ell, k)Q_a^y(\ell, k)^* = \frac{2e^{i\ell\alpha}\sin\alpha}{1 - e^{i\ell\alpha}\cos\alpha}|Q_a^y(\ell, k)|^2$$
from which we directly find
$$P_a^{xy}(\ell, k) = \frac{2\sin\alpha(\cos\ell\alpha - \cos\alpha)}{1 - 2\cos\ell\alpha\cos\alpha + (\cos\alpha)^2}|Q_a^y(\ell, k)|^2 \quad (A10)$$
which clearly becomes zero once $\ell = \pm 1$.; at the same time we have
$$S_a^z(\ell, k) = -\frac{2\sin\alpha\sin\ell\alpha}{1 - 2\cos\ell\alpha\cos\alpha + (\cos\alpha)^2}|Q_a^y(\ell, k)|^2 \quad (A11)$$
from which it is simple to obtain $|S_a^z(\ell = \pm 1, k)| = 2P_a^{yy}(\ell = \pm 1, k) = 2P_a^{xx}(\ell = \pm 1, k)$. This means that if we calculate the ratio $R_\beta$ considering only the contributions of the atoms on the column centered in (0; 0) we obtain value of 0.5, independently from the sample thickness. The deviations of $R_\beta$ from this limiting value are due to the contributions of the atoms on the neighboring columns, which are generally smaller in strength because of the weak de-localization of the inelastic processes.

**Appendix B**

The target of this section is to find a relation between $M(\Delta E)$, $N_{ii}(\Delta E)$ (integrated over the atomic edges L$_2$ and L$_3$) and the orbital and spin components of the atomic magnetic moment, starting from the equations provided in Ref.5 and summarized in the Supplementary Material (SM) (Section 1).
We define the quantities
$$M_{23} = \int\limits_{L_2 + L_3} M(\Delta E) d\Delta E$$

$$M_2 = \int_{L_2} M(\Delta E) d\Delta E$$
$$M_3 = \int_{L_3} M(\Delta E) d\Delta E$$

i.e. the integral over the edges $L_2$ and/or $L_3$ of the energy dependent imaginary part of the MDFF. If we start from Eq.1 in the SM, we have

$$M_{23} = \sum_J Im[S_J(\boldsymbol{q},\boldsymbol{q'})]\frac{1}{[\boldsymbol{q}\times\boldsymbol{q'}]_z} = \frac{1}{2}\frac{9G_L^2}{2L+1}<\boldsymbol{L}>_z$$

where the orbital component of the atomic magnetic moment is given by $m_{orb} = \mu_B <\boldsymbol{L}>_z$, being $<\boldsymbol{L}>_z$ the expectation value of the atomic OAM along $z$ axis; the definition of $G_L^2$ is given by Eq. 6 of the SM, while the index $J$ runs over both $L$ edges.

At the same time we can exploit Eq.3 of the SM to evaluate the integral of $M(\Delta E)$ over the two distinct atomic edges, i.e.

$$M_2 = \frac{1}{[\boldsymbol{q}\times\boldsymbol{q'}]_z}\int_{L_2} Im[S(\boldsymbol{q},\boldsymbol{q'},\Delta E)]d\Delta E$$
$$= \frac{9G_L^2}{(2L-1)(2L+1)}\left[\frac{L-1}{2}<\boldsymbol{L}>_z\right.$$
$$\left. - \frac{<\boldsymbol{S}>_z (L-1)L}{3}\left(1+\frac{2L+3}{L}\frac{<\boldsymbol{T}>_z}{<\boldsymbol{S}>_z}\right)\right]$$

$$M_3 = \frac{1}{[\boldsymbol{q}\times\boldsymbol{q'}]_z}\int_{L_3} Im[S(\boldsymbol{q},\boldsymbol{q'},\Delta E)]d\Delta E$$
$$= \frac{9G_L^2}{(2L-1)(2L+1)}\left[\frac{L}{2}<\boldsymbol{L}>_z\right.$$
$$\left. + \frac{<\boldsymbol{S}>_z (L-1)L}{3}\left(1+\frac{2L+3}{L}\frac{<\boldsymbol{T}>_z}{<\boldsymbol{S}>_z}\right)\right]$$

As pointed out by Chen *et al*[17]., the ratio $\frac{<T>_z}{<S>_z}$ is generally negligible in case of bulk systems, so we have neglected its contributions in the derivations presented in this work.

As the last point we need to find an expression for the functions $N_{ii}(\Delta E)$ integrated over both the edges: to do this we use Eq.2 of the SM, from which we can write

$$\sum_J Re[S_J(\boldsymbol{q},\boldsymbol{q'})] = \int_{L_2+L_3} d\Delta E[\boldsymbol{q}\cdot\mathbb{N}(\Delta E)\cdot\boldsymbol{q'}]$$
$$= \frac{9L^2 G_L^2}{(2L-1)(2L+1)}\boldsymbol{q}\cdot\boldsymbol{D}\cdot\boldsymbol{q'}$$
$$+ \sum_{i\neq j} q_i q'_j O_{ij} \quad (B1)$$

where $\boldsymbol{D}$ is a diagonal tensor such that

$$D_{ii} = \frac{2(2L+1)(2L-1)}{3L} - N_e + \frac{1}{2L^2}<\boldsymbol{L}^2>_i$$

with $i = x,y,z$, while $O_{ij}$ is a 3x3 tensor with null elements on the diagonal, which can be derived from the second addendum of the sum in Eq.(2) of the SM. By equating the coefficients of $q_x q'_x$ and $q_y q'_y$ of the second and third members of Eq.B1 we can obtain

$$\int_{L_2+L_3} N_{ii}(\Delta E)\, \Delta E = \frac{9L^2 G_L^2}{(2L-1)(2L+1)}D_{ii}$$

from which it is simple to derive the integral of the sum of $N_{xx}(\Delta E)$ and $N_{yy}(\Delta E)$ needed in the derivation of the sum rules.

We also point out that calling as $f$ the final states in which the sample electron can be excited, and exploiting the relation $<\boldsymbol{L}_x> + <\boldsymbol{L}_y> = <\boldsymbol{L}^2> - <\boldsymbol{L}_z>$ we have

$$D_{xx} + D_{yy} = \frac{4(2L+1)(2L-1)}{3L} - 2N_e$$
$$+ \sum_f \sum_{m_\ell,m_s} (\hbar^2 L(L+1)$$
$$- \hbar^2 m_\ell^2)|<m_\ell m_s|f>|^2$$

with $L = 2$ and the indexes $m_\ell$ and $m_s$ running over the intervals $\{-2,-1,0,1,2\}$ and $\left\{\frac{-1}{2},\frac{1}{2}\right\}$: this quantity can be evaluated by *ab initio* calculations.